\documentstyle[aps,twocolumn,epsfig]{revtex}
\begin{document}
\newcommand{\ve}[1]{\mbox{\boldmath $#1$}}
\twocolumn[\hsize\textwidth\columnwidth\hsize
\csname@twocolumnfalse%
\endcsname

\draft

\title{Probing Bose-Einstein Condensation of Excitons with 
       Electromagnetic Radiation}
\author{K. Johnsen and G. M. Kavoulakis}
\date{\today} 
\address{NORDITA, Blegdamsvej 17, DK-2100 Copenhagen \O, Denmark}

\maketitle

\begin{abstract}

   We examine the absorption spectrum of electromagnetic radiation
from excitons, where an exciton in the $1s$ state absorbs a photon
and makes a transition to the $2p$ state. We demonstrate that the
absorption spectrum depends strongly on the quantum degeneracy of 
the exciton gas, and that it will generally manifest many-body effects.
Based on our results we propose that absorption of infrared
radiation could resolve recent contradictory experimental
results on excitons in Cu$_2$O.

\end{abstract}
\pacs{PACS numbers: 05.30.Jp, 71.35.Lk, 71.35.-y, 71.35.Cc}

\vskip0.0pc]

   The phenomenon of Bose-Einstein condensation has 
attracted much attention in the recent years, and many
experimental groups have reported the
formation of Bose-Einstein condensates in vapors of
alkali-metal atoms \cite{RMP}. Excitons, bound states of electrons
and holes in semiconducting materials,
are another candidate for undergoing this phase transition \cite{Knox}.
Since excitons are composite particles consisting of two
fermions, they are expected to obey Bose-Einstein statistics in the
limit where their thermal de Broglie wavelength becomes comparable 
to their interparticle spacing, provided that this spacing
is much larger than the exciton Bohr radius. 

   Excitons in Cu$_2$O, which is a dipole forbidden material,
have some important advantages in this respect \cite{MS}. Their
radiative lifetime is long and their binding energy
is much larger than the typical thermal energies. In addition 
the effective electron and hole 
masses are isotropic, and finally there are no bound states
between excitons, i.e., biexcitons, or an electron-hole-liquid phase.

    A lot of effort has been made in order to create a Bose-Einstein
condensate of excitons in Cu$_2$O \cite{Andre,Jim}. To determine the density
and the temperature in the above experiments, the phonon-assisted
recombination spectrum was fitted to a Bose-Einstein
distribution, which gave the chemical potential and the
temperature -- two essentially independent parameters.
Given the total exciton mass, the density was then evaluated
to be on the order of $10^{18}$ cm$^{-3}$, while the temperature
was on the order of 20 -- 30 K, higher than the lattice temperature which
was kept at about 5 K. Therefore, according to this approach, the exciton
gas was very close to the phase boundary for Bose-Einstein condensation,
and the angular-momentum singlet-state (para)excitons were the species
reported to actually cross the phase boundary.

   More recent experiments \cite{Keith} have, however, questioned
the older method of determining the density and the temperature.
In these experiments the number of excitons was determined directly
and, with a relatively reliable estimate of the volume of the exciton cloud, 
the density was found to be two to three orders of magnitude lower,
i.e., around $10^{16}$ cm$^{-3}$, where the exciton gas should show
no sign of quantum degeneracy.

  One, therefore, needs to find a reliable method of determining the 
density and in particular the degree of quantum degeneracy of excitons.
In this study we propose that measuring the absorption spectrum of
infrared radiation which induces transitions of the excitons from the
$1s$ to the $2p$ state can indeed resolve the discrepancy.
Our study is also directly applicable to other systems, like 
excitons in quantum wells, and thus it could help resolve 
some other experimental observations which are controversial \cite{QW}.

   \"Oktel and Levitov \cite{OL} have examined a similar problem as the one
we consider here, in the context of optical excitations of hydrogen atoms 
and have studied the many-body effects that show up in the absorption 
spectrum, for an effective contact potential between the atoms. Our 
approach is equivalent to theirs in the limit of equal masses
between the excitons in the $1s$ and the $2p$ state. In another 
study Pethick and Stoof \cite{PS} have considered a more general 
form of the interatomic potential. Finally Lewenstein and You 
\cite{LY} have examined the enhancement of 
scattering of light from a gas of bosonic atoms as they form
a Bose-Einstein condensate, and have suggested that this effect
could be used for the detection of the condensate. 

   This Letter is organized as follows: We first consider the
case of an ideal exciton gas, and examine the relevant energy scales
that enter the problem and also derive simple expressions for the
absorption spectrum. We then examine the problem of
an interacting exciton gas within the Hartree-Fock approximation
and find that the
interactions can have a very drastic effect on the absorption spectrum.  
Finally we present our results with the interactions included, and
show that infrared absorption can be used in order to determine the
degree of quantum degeneracy of excitons, thus proposing an experiment
which could resolve this issue.

   Consider the process in which
an exciton in the $1s$ state, with momentum $\hbar \bf k$,
absorbs a photon of momentum $\hbar \bf q$, making a transition to the $2p$ 
state with momentum $\hbar({\bf k}+{\bf q})$. The conservation of energy 
in this process implies that
\begin{eqnarray}
   \epsilon_{\bf k}^{1s} + \hbar c q = \epsilon_{{\bf k}+{\bf q}}^{2p},
\label{consene}
\end{eqnarray}
where $\epsilon_{\bf k}^{i} = E_{i} + {\hbar^2 k^2}/{2 m_{i}}$
with $E_{i}$ being the binding energy of the $i$ state, and
$m_{i}$ being the total exciton mass in the state $i$. 
In Cu$_2$O $m_{1s} \approx 3m$, where 
$m$ is the electron mass, is larger than the sum of the effective 
electron and hole masses as a result of the small 
Bohr radius of the $1s$ excitons, $a_B^{1s} \approx 5.3$ \AA\, 
compared to the lattice constant $a_l \approx 4.26$ \AA \cite{KCB}.
On the other hand, the Bohr radius of excitons in the $2p$ state $a_B^{2p}$
is given by the hydrogenic formula which yields $\approx 44$ \AA\
\cite{KCB}. Since $a_B^{2p} \gg a_l$, $m_{2p}$ is 
expected to be equal to the sum
of the effective electron and hole masses, which is $\approx 1.68m$.
In Eq.\,(\ref{consene}) there are two distinct energy scales, i.e., the
energy separation $\Delta E=E_{2p}-E_{1s} \approx 55$ meV, and the thermal
energy $\hbar^2 k^2/2m \sim k_B T$, which is of order 1--10 meV. Since 
$\hbar q c \sim \Delta E$ and $\hbar^2 k^2/2m \sim k_B T$, 
we get that $q/k \sim \Delta E/\sqrt{m c^2 k_B T} \sim 10^{-3}$. 
Therefore $\hbar^2 q k /2m \sim 10^{-3} k_B T$, and $\hbar^2 q^2/2m 
\sim 10^{-6} k_B T$, which allows us to neglect the corresponding terms
in Eq.\,(\ref{consene}). Solving in terms of $k^2(q)$, we obtain
\begin{eqnarray}
	k^2(q) \approx \frac {2 m_{1s} m_{2p}} {m_{1s}-m_{2p}}
   \, \frac {\hbar c q - \Delta E} {\hbar^2},
\label{kq}
\end{eqnarray}
which gives the magnitude of the momentum $\hbar {\bf k}$ of
the exciton in the $1s$ state that absorbs a photon with wavevector $\bf q$
and is excited to the $2p$ state. In this approximation, for 
$\hbar c q = \Delta E$ only the excitons with $k=0$ can participate
in the process; however for a Bose-Einstein condensate there is a
macroscopic number of excitons with $k=0$, and therefore the absorption 
spectrum has a pronounced peak, with a strong temperature dependence.
To see this more clearly, let us calculate the rate of this dipole-allowed
process of absorption of a photon. With the approximate
expression for the conservation of energy of Eq.\,(\ref{kq}),
the rate $\Gamma_T$ of non-condensed excitons in the $1s$ state
absorbing a photon and making the transition to the $2p$ state is given by
\begin{eqnarray}
   \Gamma_T = \frac {2 \pi} {\hbar} 
  \sum_{{\bf k}} |M_{\bf q}|^2
 n_{\bf k}^{1s} (1+n_{\bf k+q}^{2p}) f_{\bf q} 
\, \delta(\hbar q c - \Delta E_{{\bf k}}),
\label{gr}
\end{eqnarray}
where $M_{\bf q}$ is the matrix element of this process, 
$n_{\bf k}^i$ is the distribution function of species $i$
($1s$ or $2p$ excitons), $f_{\bf q}$ is the distribution function
of the incoming photons, and $\Delta E_{{\bf k}} = \epsilon_{{\bf k}}^{2p} -
\epsilon_{{\bf k}}^{1s}$. Neglecting the occupation number of excitons
in the $2p$ state, $n_{\bf k+q}^{2p} \ll 1$, for monochromatic radiation
Eq.\,(\ref{gr}) implies that
\begin{eqnarray}
  \Gamma_T 
    \propto (\hbar c q - \Delta E)^{1/2}
   n_{{\bf k}_0}^{1s} \, \theta(\hbar c q - \Delta E),
\label{grmod}
\end{eqnarray}
where $\theta(x)$ is the Heaviside step function, and the magnitude 
of ${\bf k}_0$ is given by Eq.\,(\ref{kq}).
The above result expresses the fact that the absorption spectrum is
proportional to the density of states times the distribution function
calculated at a wavevector with a magnitude given by Eq.\,(\ref{kq}).

   For a Bose-Einstein condensed exciton gas with $N_C$ excitons
occupying the ${\bf k}=0$ state, the rate $\Gamma_C$ of the same 
process is simply
\begin{eqnarray}
   \Gamma_C = \frac {2 \pi} {\hbar}
   |M_{\bf q}|^2
 N_C (1+n_{\bf q}^{2p}) f_{\bf q} 
\, \delta(\hbar q c - \Delta E), 
\label{grcon}
\end{eqnarray}
or $\Gamma_C \propto N_C \, \delta(\hbar c q - \Delta E)$. Therefore
the absorption spectrum (which is proportional to the decay rate) 
of an ideal Bose-Einstein condensed gas has a strong peak with a height
that scales as $N_C$. However, as we show below,
the interactions can modify this picture drastically.

   We thus turn to the more realistic problem of an interacting
Bose gas. We start with the Hamiltonian $H$ \cite{Chris}
\begin{eqnarray}
   H = \sum_{\bf k} \epsilon_{\bf k}^{1s} a_{\bf k}^{\dagger} a_{\bf k}
  + \frac {U_{11}} {2V} \sum_{\bf k, \bf k', \bf q}
 a_{\bf k+q}^{\dagger} a_{\bf k' - q}^{\dagger} a_{\bf k'} a_{\bf k}
\nonumber \\
 + \sum_{\bf k} \epsilon_{\bf k}^{2p} b_{\bf k}^{\dagger} b_{\bf k}
 + \frac {U_{12}} {V} \sum_{\bf k, \bf k', \bf q}
 b_{\bf k+q}^{\dagger} a_{\bf k' - q}^{\dagger} a_{\bf k'} b_{\bf k},
\label{hamiltonian}
\end{eqnarray}
where $V$ is the volume of the gas. In the
above Hamiltonian we assume that the excitons interact
with an effective contact potential, with
$U_{ij} = 2 \pi \hbar^2 a_{ij}/\mu_{ij}$ being the strength
of the effective two-body interaction. Here $a_{ij}$ is the scattering
length for collisions between excitons in the 
states $i$ and $j$ (1 for the $1s$ state, and 2 for the $2p$ state).
The reduced mass $\mu_{ij}$ entering the above expression is given by 
$\mu_{ij} = m_{i} m_{j} / (m_{i} + m_{j})$. Also
$a_{\bf k} (b_{\bf k})$ and $a_{\bf k}^{\dagger} (b_{\bf k}^{\dagger})$ 
are annihilation and creation operators for an exciton with momentum $\bf k$
in the $1s (2p)$ state.

   Let us now consider the ground state of the system with $N$ excitons,
which we denote as $| 0 \rangle
=|N_C, N_{{\bf k}_1}, N_{{\bf k}_2}, \dots,
N_{{\bf k}_e}, \dots \rangle$, where $N_{{\bf k}_i}$ is the
occupancy of a state with momentum ${\bf k}_i$.
Initially we take all the excitons to be in the $1s$ state. Since we
consider a Bose gas both in the normal, as well as in the condensed regime,
we assume that there is one state that can get populated by a
macroscopic number of excitons, $N_C$, and thus $N_C$ can get as high as the
total number of excitons, $N$, whereas the $N_{{\bf k}_i}$ are of order 
unity. 

   We now examine such a system when one creates excitations
of the excitons from the $1s$ to the $2p$ state with the action of some laser
pulse. If an exciton with momentum ${\bf k}_e$ is excited to the $2p$
state with momentum ${\bf k}_e^{'}= {\bf k}_e + {\bf q}$, where
${\bf q}$ is the wavevector of the laser light, since ${\bf q}$
is very small, we shall assume that we have vertical transitions, i.e.,
${\bf q} = 0$. We denote the excited states as
$|\Phi_{{\rm exc},{\bf k}_e} \rangle=|{\bf k}_e;
N_C, N_{{\bf k}_1}, N_{{\bf k}_2}, \dots,
N_{{\bf k}_e}-1, \dots \rangle$, which are the basis vectors of
our problem. The number of such states is $N-N_C+1 = N_T + 1$, where 
$N_T = N - N_C$ is the number of $1s$ excitons in states with ${\bf k} 
\neq 0$. The laser beam that excites the excitons from the $1s$ to the
$2p$ state creates a superposition of the states 
$|\Phi_{{\rm exc},{\bf k}_e} \rangle$ \cite{Chris}. 
Thus, in order to determine 
the absorption spectrum, we consider the matrix with elements
$H_{i,j} = \langle \Phi_{{\rm exc},{\bf k}_i}|H|\Phi_{{\rm exc},{\bf k}_j}
\rangle - \langle 0 | H | 0 \rangle$. One finds that
\begin{eqnarray}
   H_{i,j} = \delta_{i,j} [ \Delta E_{{\bf k}_i}
   + U_{11} (n_{{\bf k}_i} - 2n)
 &+& U_{12} (n - n_{{\bf k}_i})]
\nonumber \\
  &+& U_{12}\sqrt{n_{{\bf k}_i} n_{{\bf k}_j} },
\label{me}
\end{eqnarray}
where $n=N/V$ and $n_{{\bf k}_i} =n_{{\bf k}_i}^{1s}= N_{{\bf k}_i}/V$ 
is the Bose-Einstein distribution for the $1s$ excitons.
Let $\Psi_i$ be the components of an eigenvector with 
eigenvalue $E$. Starting from the eigenvalue equation 
$\sum_{j=0}^{N_T} H_{i,j} \Psi_j = E \Psi_i$,
we solve in terms of $\Psi_i$, multiply by
$\sqrt{n_{{\bf k}_i}}$ and sum over $i$. Eliminating
the factor $\sum_{j=0}^{N_T} \Psi_j \sqrt{n_{{\bf k}_j}}$ 
from the resulting equation, 
the eigenvalues of $H_{i,j}$ are then given by the roots
of $g(E)-1=0$, where
\begin{equation}
   g(E) = \sum_{i=0}^{N_T} \frac {U_{12} n_{{\bf k}_i}}
  { E - [\Delta E_{{\bf k}_i} + U_{11} 
 (n_{{\bf k}_i} - 2 n) + U_{12} (n - n_{{\bf k}_i})]}.
\label{eigequ}
\end{equation}
Distinguishing the condensed state $(i=0)$ from the 
other states $(i \neq 0)$, Eq.\,(\ref{eigequ}) takes 
the following form in the thermodynamic limit
\begin{eqnarray}
   g(E) =
 \frac {U_{12} n_c}{E - [\Delta E + U_{11} (n_c - 2 n) + U_{12} (n - n_c)]}
\nonumber \\
    + \sum_{i \neq 0}^{N_T} \frac {U_{12} n_{{\bf k}_i}}
  { E - [\Delta E_{{\bf k}_i} + n (U_{12} - 2 U_{11})]} ,
\label{eigequn}
\end{eqnarray}
where $n_C = N_C/N$.
In the above equation there are in general three limiting cases, 
depending on the ratio of the interaction energy $n U_{12}$,
to the typical kinetic energy $\Delta E_{{\bf k}_i}$, which is 
on the order of the thermal energy, $k_B T$.
In the limit $n U_{12} \ll k_B T$, one recovers the results we
found earlier for the ideal Bose gas. In the opposite limit,
$n U_{12} \gg k_B T$, the behaviour of the system of excitons
is ``collective". A graphical solution of the eigenvalue equation 
shows that in the condensed phase, where both $N_C$ and $N_T$ are
of order $N$, there are two strong modes, which give rise to
two peaks in the absorption spectrum. There are 
also $N_T-1$ solutions, which form a continuum corresponding to
single-particle excitations of the thermal excitons. In the same limit
$n U_{12} \gg k_B T$ for a fully condensed gas
as well as for a gas in the normal state, there is only one mode,
since then one has a one-component system. Finally when 
$n U_{12} \sim k_B T$ the system behaves in a ``mixed" way.
In addition, the limit
$|m_{1s}/m_{2p}-1| \ll 1$, is equivalent to the case $n U_{12} \gg k_B T$
and Eq.\,(\ref{eigequn}) reduces to a quadratic algebraic equation,
which gives the same result as the one derived by \"Oktel and Levitov
in Ref.\,\cite{OL}. 

   By adding a small imaginary part in $g(E)$, i.e., $g(E+ i \eta)$,
where finite $\eta$ results in homogeneous broadening of the energy levels, 
we calculate the corresponding imaginary part of the susceptibility
$[g(E+ i \eta)-1]^{-1}$
obtaining the absorption spectra shown in Fig.\,1. Broadening can be
calculated from first principles \cite{PS} -- however in the present study 
we assume small homogeneous broadening, choosing $\eta = 10^{-2}$ meV to 
produce the graphs in Fig.\,1. The broadening of the energy levels is
expected to be small, and this can be seen by
examining the three basic mechanisms which contribute to that, i.e.,
the exciton-exciton elastic collisions, their 
scattering with the lattice, and their radiative lifetime. 
The radiative lifetime of the orthoexcitons in the $1s$ state 
$\tau_o^{1s}$ is $\sim 10^{-5}$s \cite{Keith} and that of the paraexcitons
$\tau_p^{1s}$ is $\sim 10^{-3}$s. The radiative lifetime in the $2p$ state
$\tau_i^{2p}$ is smaller by a factor of $(k_{\gamma} a_B^{2p})^{2}$,
since the transition is dipole allowed,
where $k_{\gamma}$ is the wavevector of the emitted photon.
Since $k_{\gamma} = E_g / \hbar c$, where $E_g \approx 2.17$ eV
is the gap energy, $k_{\gamma} \approx 10^{-3}$ \AA$^{-1}$. 
Therefore $\tau_o^{2p} \sim 10^{-8}$ s, while
$\tau_p^{2p} \sim 10^{-6}$ s. The exciton-phonon scattering time is 
on the order of $10^{-9}$s \cite{KBW}. Finally for a density as high as
$10^{18}$ cm$^{-3}$, the exciton-exciton scattering time is on the 
order of 10$^{-11}$s \cite{KM}, which turns out to be the shortest possible 
timescale, giving an energy broadening of less than 0.1 meV.

   We now analyze the results shown in Fig.\,1.
To produce these graphs, we made use of the results of Ref.\,\cite{Ceperley},
that $a_{11} = 2.1 a_B^{1s}$ for paraexcitons, and assumed that 
$a_{11} = 10$ \AA. For the value of
$a_{12}$ very little is known and for this reason we have considered
both the case of positive (left column in Fig.\,1), as well 
as negative (right column in Fig.\,1) values. The ratio
$|a_{12}/a_{11}|$ is expected to be larger than 1, since 
$a_B^{2p} = 44$ \AA, which is much larger than $a_B^{1s} 
\approx 5$ \AA. We made the conservative choice 
$|a_{12}/a_{11}| = 2$, although this ratio could be
larger. We also considered a temperature of 10 K for
the exciton gas in all the cases, and we varied the density
from $10^{16}$ cm$^{-3}$ to $5 \times 10^{18}$ cm$^{-3}$.
With these values $k_B T \sim 1$ meV, while $|n U_{12}|$ is 
$\sim 10^{-2}$ meV for $n=10^{16}$ cm$^{-3}$, and 
$\sim 5$ meV for $n=5 \times 10^{18}$ cm$^{-3}$. Figures 1(a+) and (a--) show 
a completely classical gas, and since $n U_{12} \ll k_B T$, 
the system behaves like an ideal gas. In Figs.\,1(b+) and (b--)
the excitons are essentially on the phase boundary for condensation
and since $n U_{12} \sim k_B T$, the system is in the ``mixed" state
where both collective and single-particle-like behaviours
show up. Figure 1(b--) shows these two distinct types of excitation, 
while Fig.\,1(b+) does not, because the collective mode is buried inside 
the continuum. In Figs.\,1(c+) and (c--) the excitons are in the condensed 
phase with $N_C/N \approx 0.48$. This is the source of the sudden  
appearance of the peak in Fig.\,1(c+). In Fig.\,1(c--) in addition to 
the two peaks, there is a contribution from the continuum that is hardly
visible. Finally in Figs.\,1(d+) and (d--) $N_C/N \approx 0.79$, and 
since $n U_{12}$ is about $5 k_B T$, the spectrum is dominated by the
collective behaviour, as the two peaks indicate. However, we remark
that the Hartree-Fock approximation does not capture effects due to 
condensate fluctuations which may be relevant in the regime
$n U_{12} \gg k_B T$.

   Let us now examine the possible experiment which could be performed
in order for these effects to be explored.
The energy of the absorbed radiation would have to be in the infrared,
with an energy of order $\Delta E \approx 55$ meV. The corresponding
wavelength is about 20 $\mu$m, and it is comparable to the size of the
cloud. Free-electron lasers provide tunable radiation in this regime.
It is important to mention that at such low energies the
crystal is transparent and the absorption of radiation due to the
process we study should be the dominant mechanism. Our analysis
requires that the infrared pulse should be sufficiently long, 
so that its energy spread is much less  
\noindent
%\eject
\begin{figure}
\begin{center}
\epsfig{file=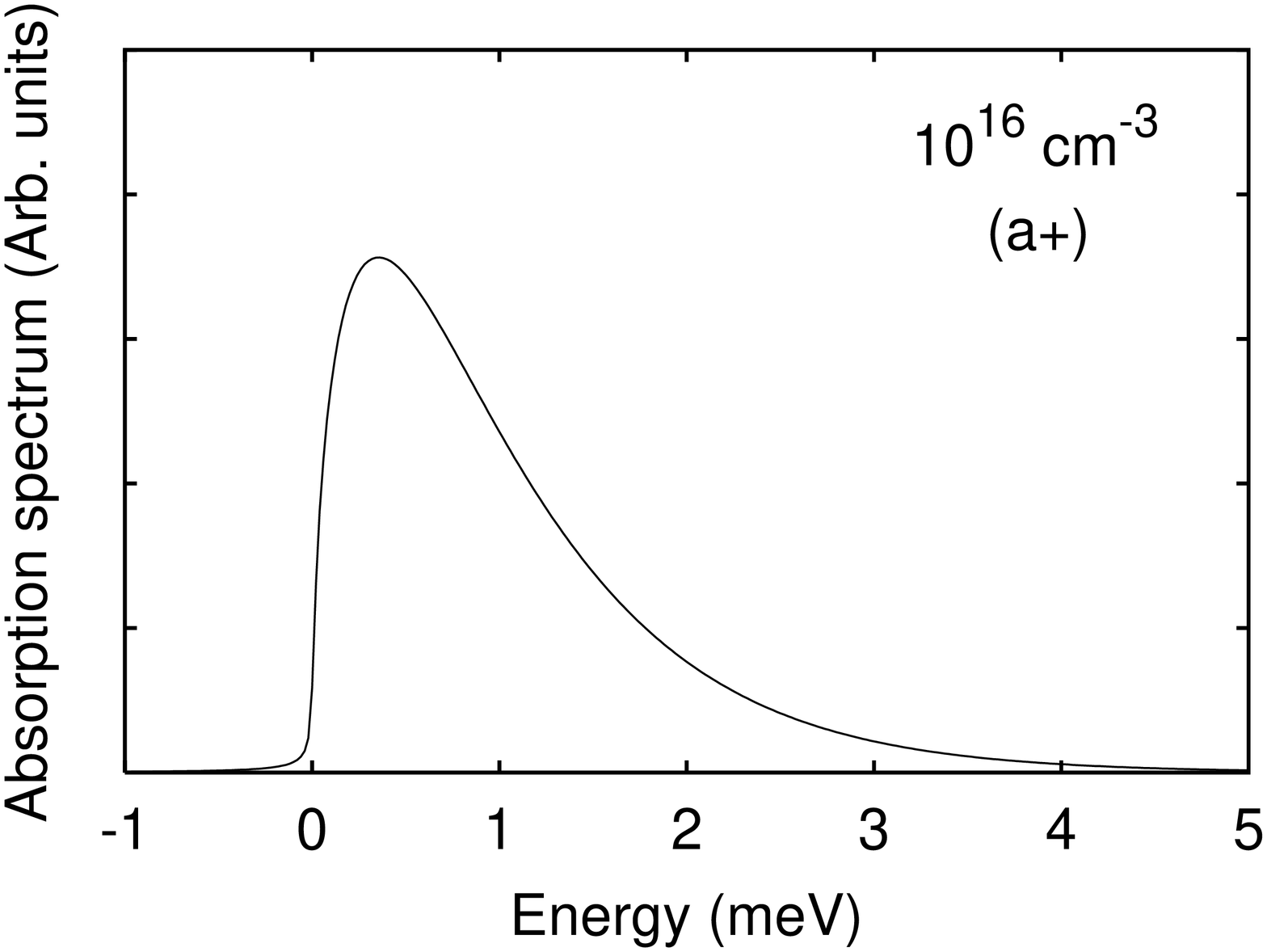,width=3.3cm,height=4.2cm,angle=-90}
\epsfig{file=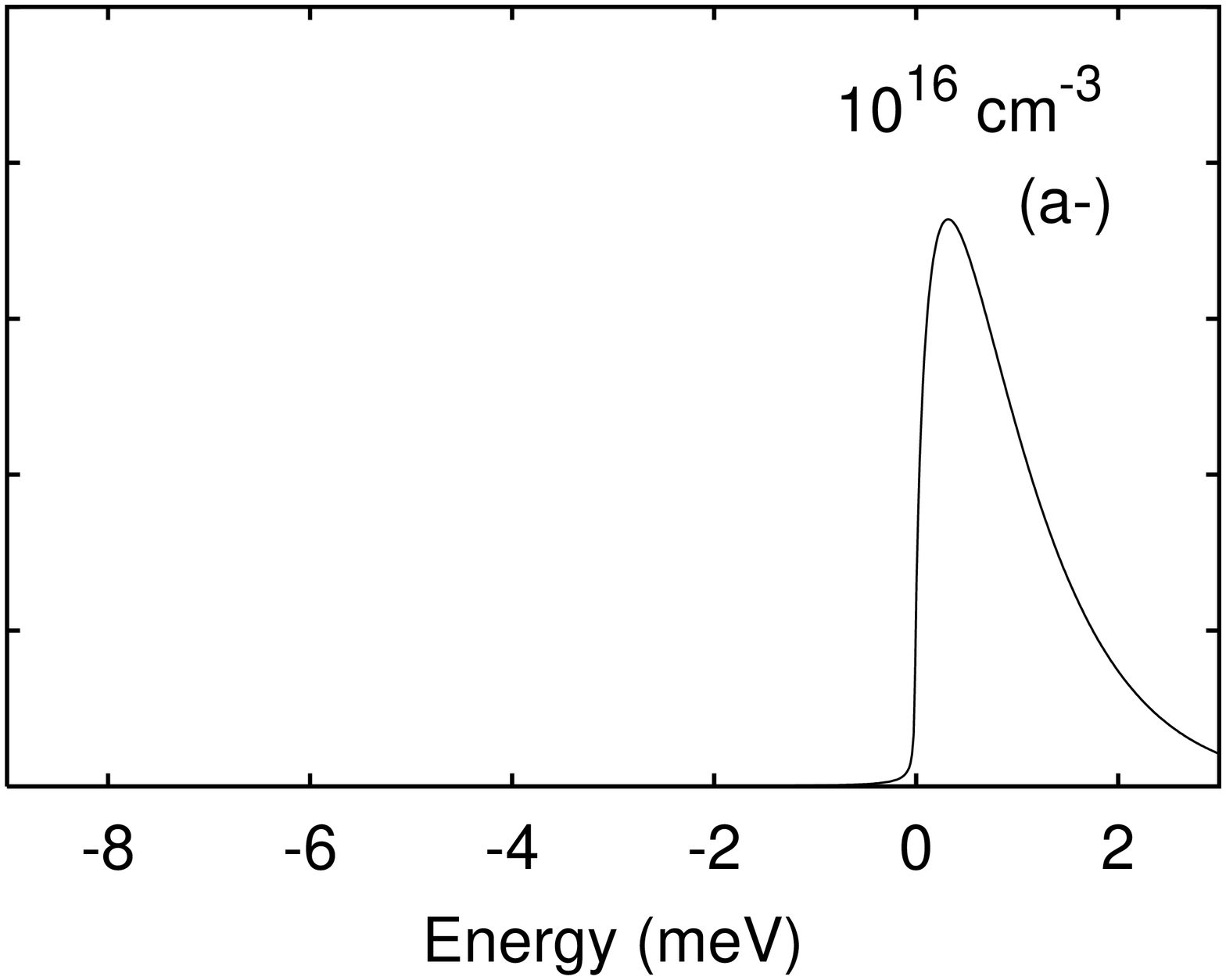,width=3.3cm,height=4.2cm,angle=-90}
\epsfig{file=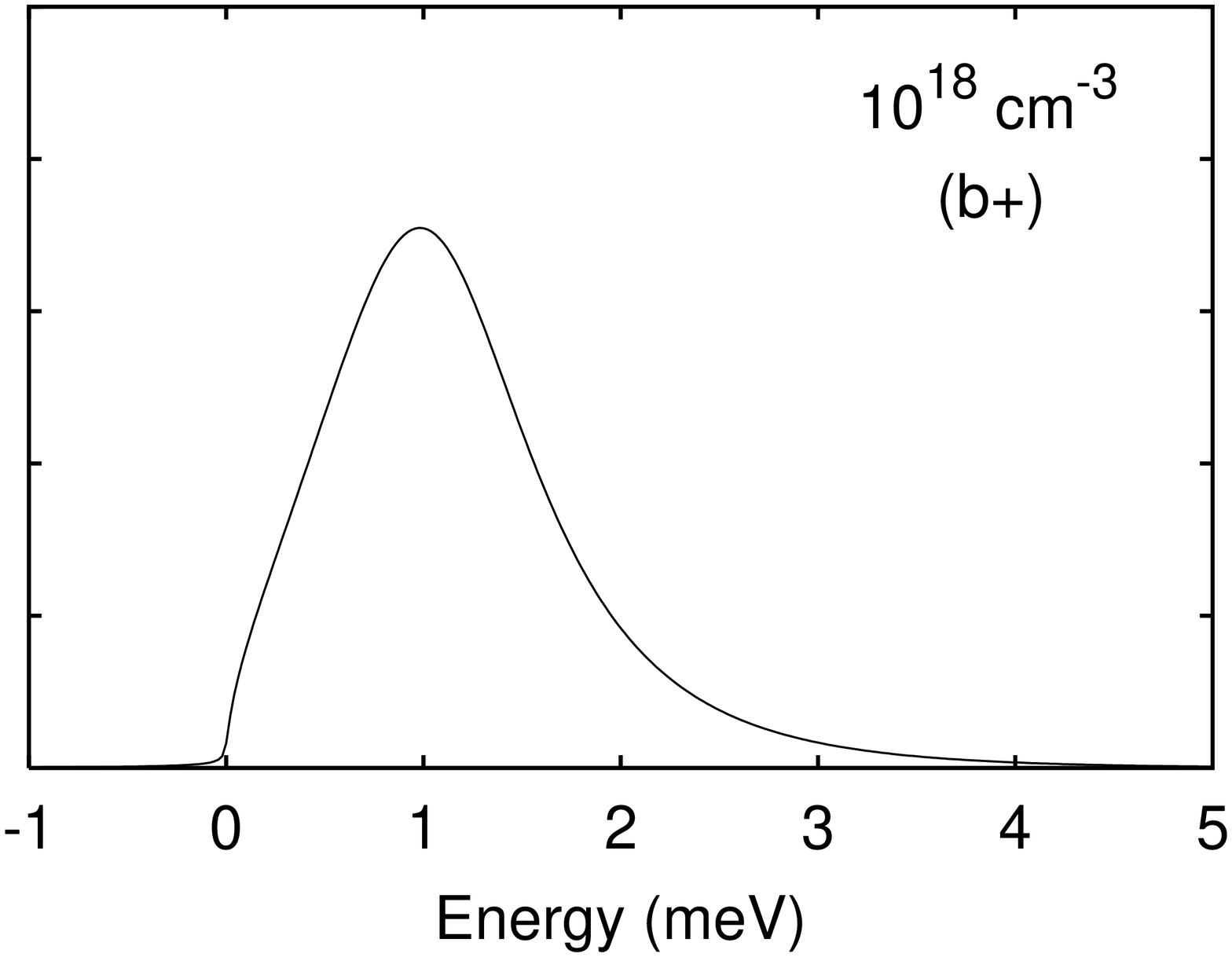,width=3.3cm,height=4.2cm,angle=-90}
\epsfig{file=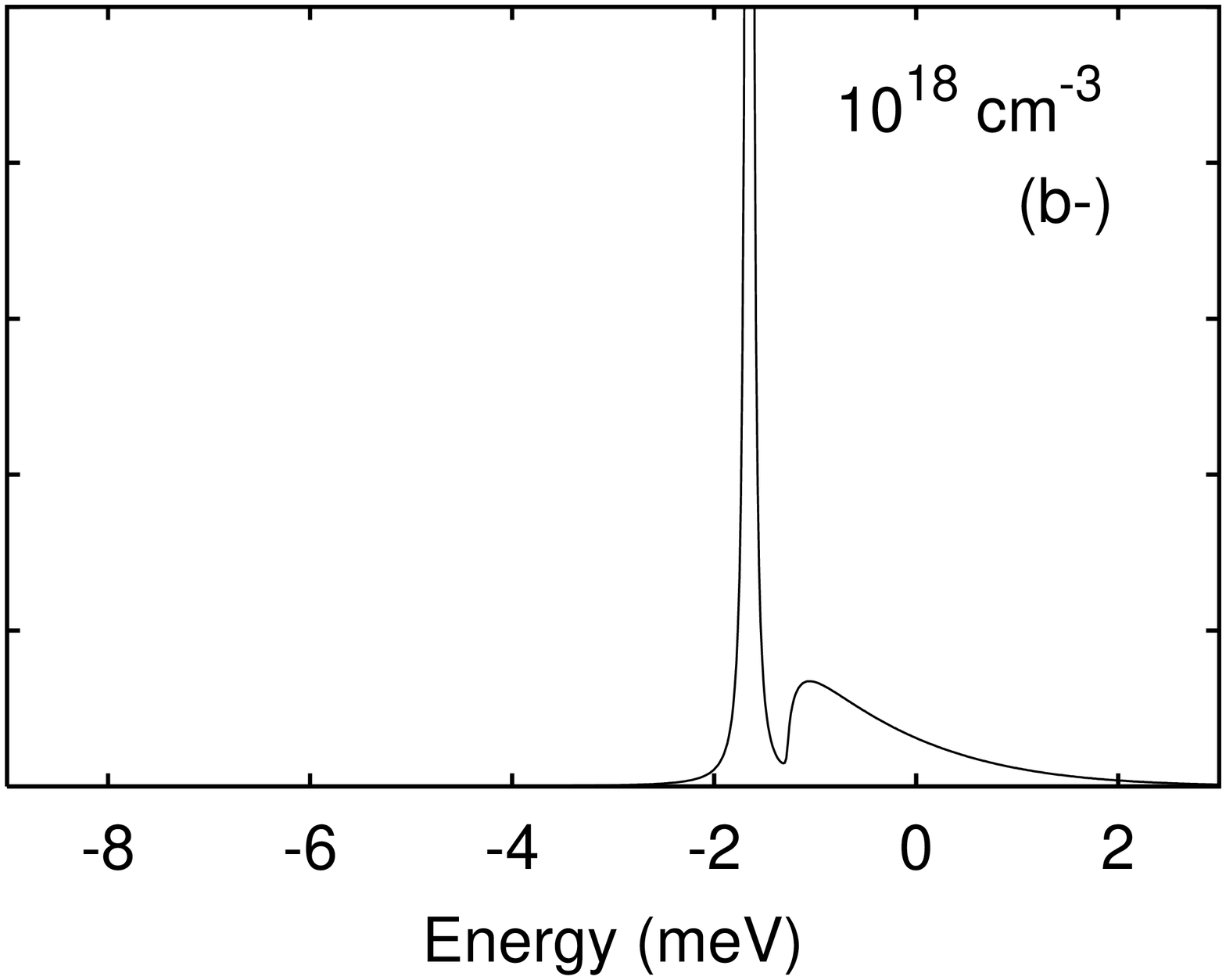,width=3.3cm,height=4.2cm,angle=-90}
\epsfig{file=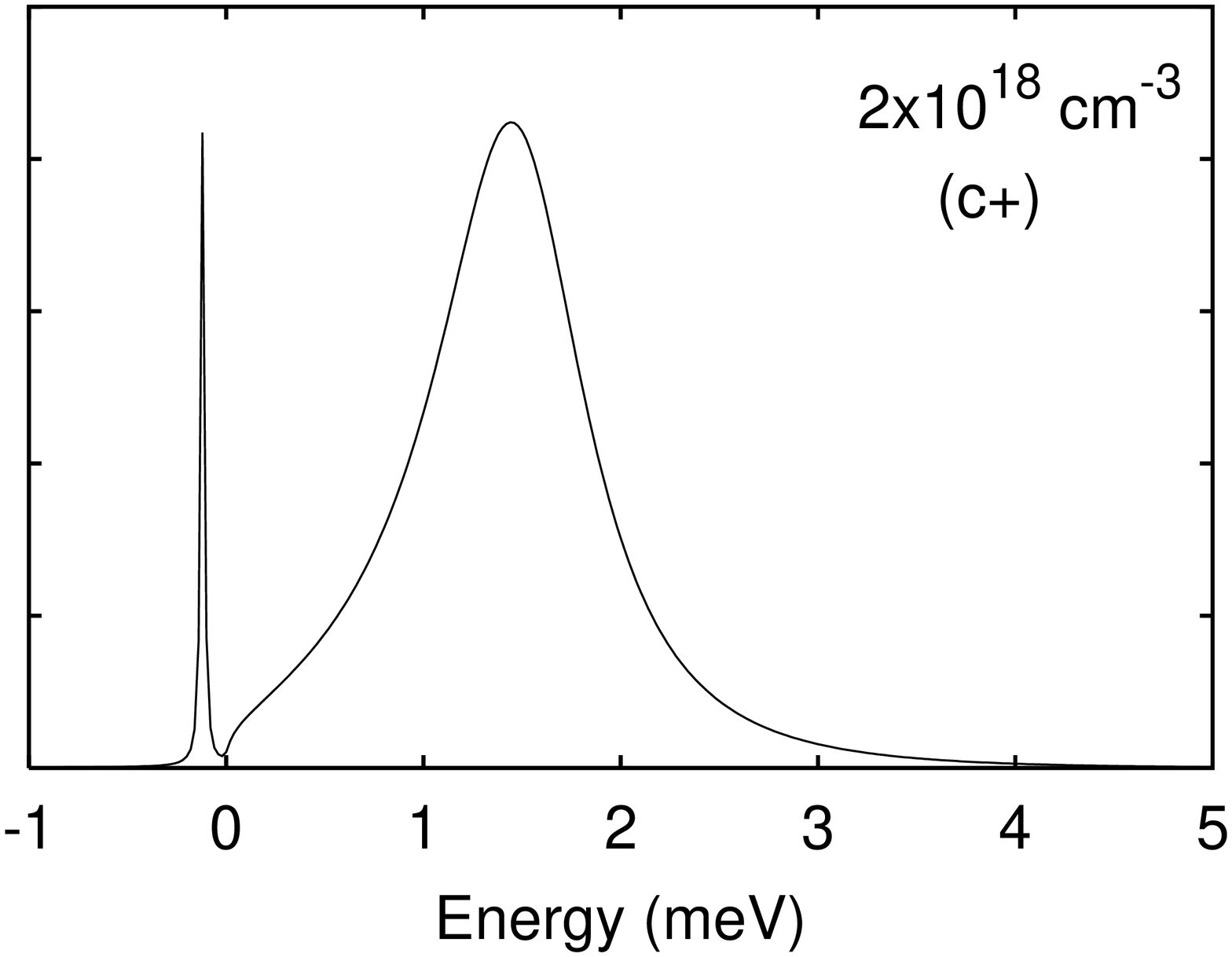,width=3.3cm,height=4.2cm,angle=-90}
\epsfig{file=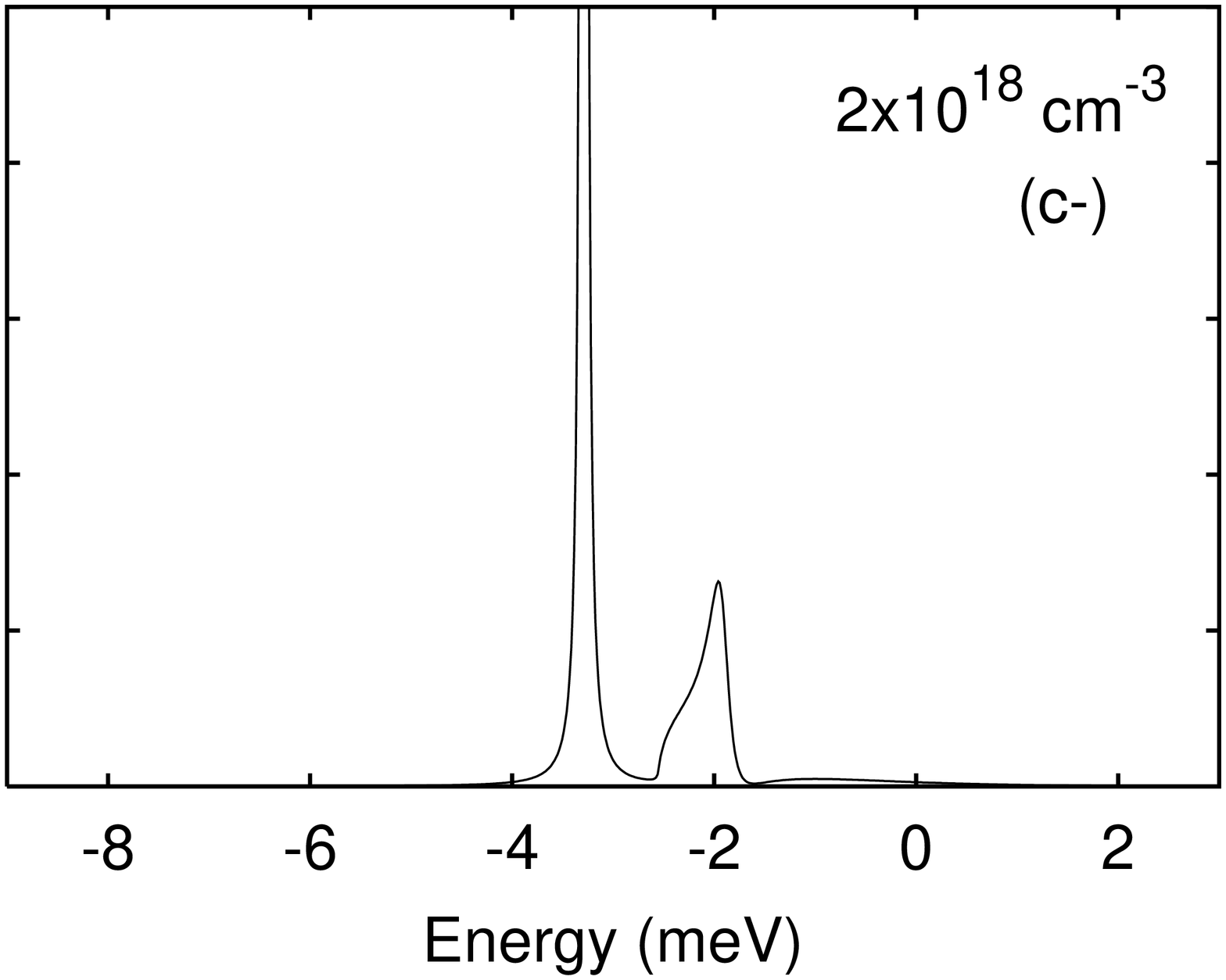,width=3.3cm,height=4.2cm,angle=-90}
\epsfig{file=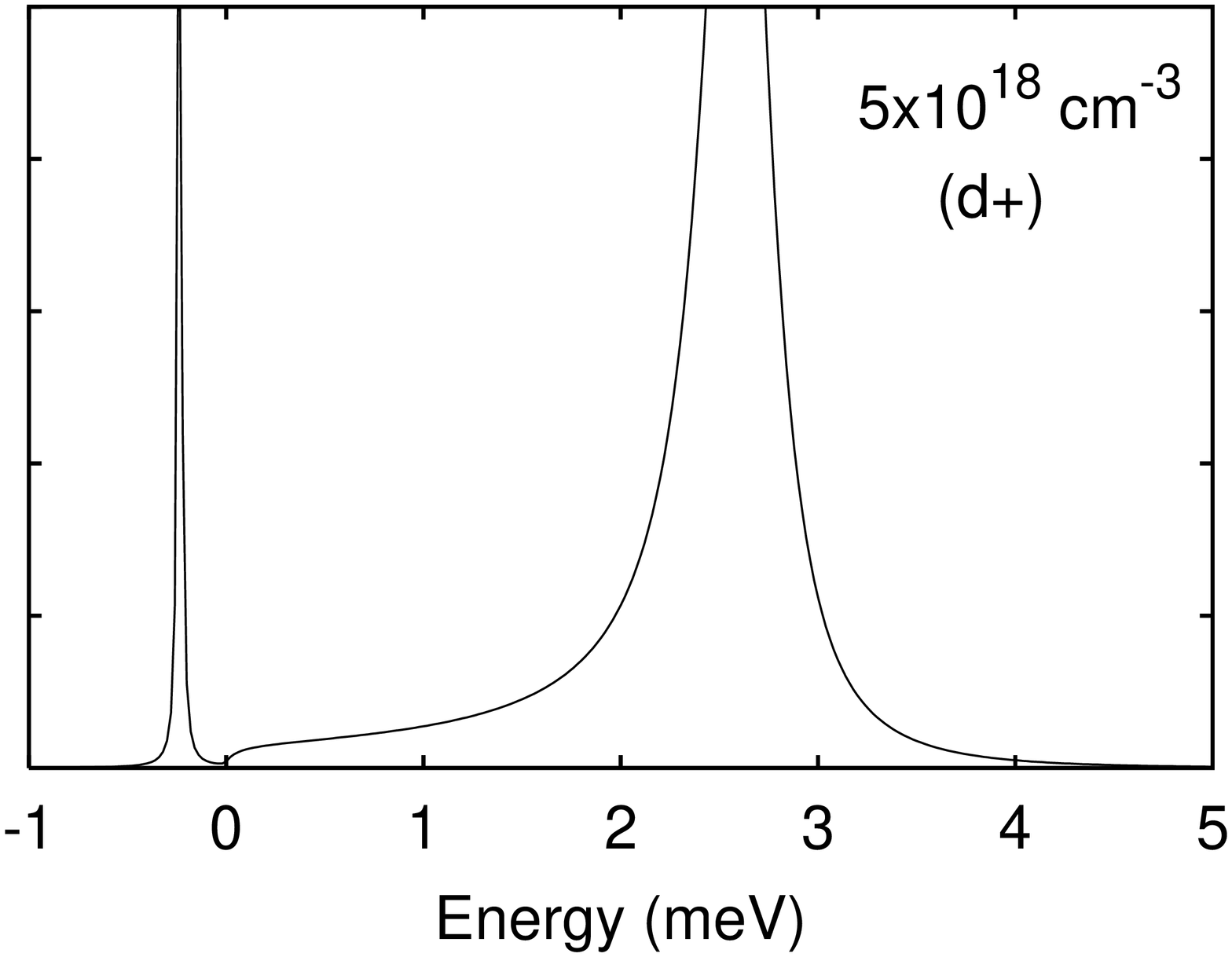,width=3.3cm,height=4.2cm,angle=-90}
\epsfig{file=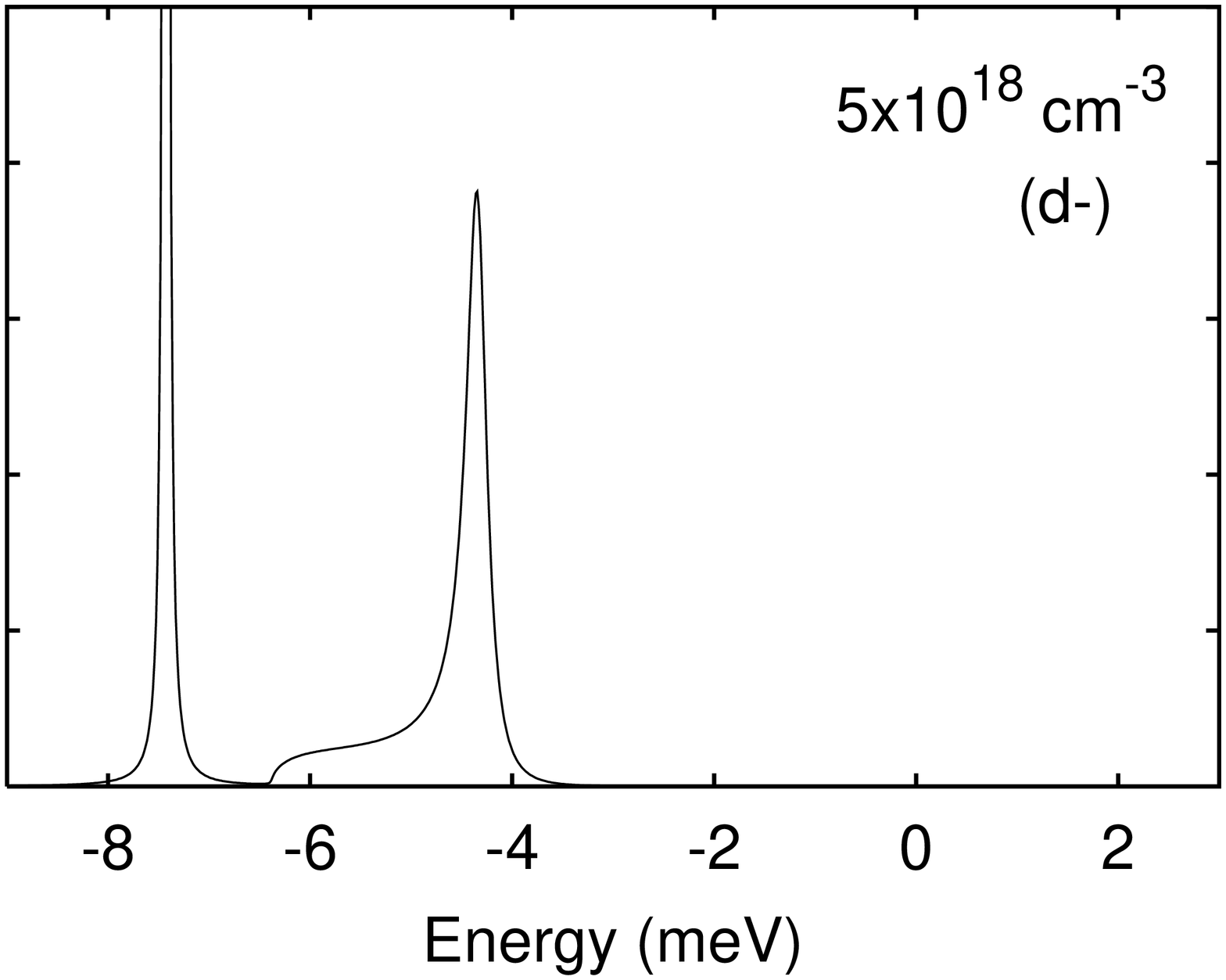,width=3.3cm,height=4.2cm,angle=-90}
\vskip 0.5pc
\begin{caption}
{The absorption spectrum on a linear scale, as a
function of the energy of the photon that is absorbed. The
absorption is in arbitrary units, but the scale is the same in
all the figures. The
energy is measured in units of meV, and its zero is measured with respect
to $\Delta E$. The scattering length $a_{11}= 10$ \AA, the
exciton temperature is 10 K, and the density is $10^{16}$, $10^{18}$,
$2 \times 10^{18}$, and $5 \times 10^{18}$ cm$^{-3}$ from
top to bottom. For the graphs on the left $a_{12}=20$ \AA,
and for the ones on the right, $a_{12}=-20$ \AA. If $\mu$ is the 
chemical potential of the gas,
$-\mu/k_B T=3.7$ in (a$\pm$), and $5 \times 10^{-4}$ in (b$\pm$), 
while $N_C/N=0.48$ in (c$\pm$), and 0.79 in (d$\pm$).}
\end{caption}
\end{center}
\label{FIG1}
\end{figure}
\noindent
than the energy width of the structures shown in Fig.\,1.
An advantage of the method we suggest is that it provides
an independent method of probing the kinetic energy
distribution of excitons. The difference between the uppermost
graphs in Fig.\,1 and the lowest is pronounced, and one should
be able to distinguish clearly the degree
of degeneracy of the excitons. In addition, this
method does not depend on the strength of the phonon-assisted
recombination line of paraexcitons, which is very weak, and since it
is close to other much stronger lines, observing this line is very hard
\cite{Andre,Jim,Keith}.

  In conclusion, we have demonstrated that the absorption spectrum
of electromagnetic radiation which induces transitions of the excitons 
from the $1s$ to the $2p$ state is strongly affected by the quantum 
degeneracy of the exciton gas and by many-body effects. 
We have thus concluded
that the absorption spectrum can provide a powerful tool for
resolving recent experimental contradictory results on excitons
in Cu$_2$O, and in quantum wells. 

\vskip0.5pc

  We are grateful to Chris Pethick and Andy Jackson for many useful 
discussions, and for their crucial input. 
K.J. thanks the hospitality of The Erwing Schr\"{o}dinger International
Institute for Mathematical Physics (Vienna).
G.M.K. was supported by the European Commission, 
TMR program, contract No.\,ERBFMBICT 983142. 
G.M.K.  would like to thank the
Foundation of Research and Technology, Hellas (FORTH) for its hospitality.

\end{document}